\documentclass[a4paper,twocolumn]{aa}
\usepackage{graphicx}
\usepackage{txfonts}
\newcommand{\besancon}{Besan\c{c}on}
\newcommand{\vdag}{^\dagger}
\begin{document}

\title{The Proper Motion of the Magellanic Clouds: \\The UCAC2-Hipparcos Inconsistency}

\author{Y. Momany\inst{1} \& S. Zaggia\inst{2} 
 }
  
\offprints{Y. Momany}

\institute {Dipartimento di Astronomia, Universit\`a di  Padova,   
Vicolo dell'Osservatorio 2, I-35122 Padova, Italy%\\
%\email{momany@pd.astro.it }
\and
INAF - Osservatrio Astronomico di Trieste, via Tiepolo 11, 
I-34131 Trieste, Italy% \\ 
%\email{zaggia@ts.astro.it}
}

\date{Accepted April, 2005}
%%%%%%%%%%%%%%%%%%%%%%%%%%%%%%%%%%%%%%%%%%%%%%%%A

\abstract{
Using the USNO CCD Astrograph all-sky  Catalog (UCAC2), we measure the
mean proper motion of the two Magellanic Clouds.
Appropriately-selected  LMC     populations  show  a   proper   motion
$\langle\mu\alpha,\mu\delta\rangle\simeq(+0.84,+4.32$)     that     is
significantly  higher, in  $\langle\mu\delta\rangle$,  than  currently
accepted                    Hipparcos-like                     values;
$\langle\mu\alpha,\mu\delta\rangle\simeq(+1.94,-0.14)$.
A higher $\langle\mu\alpha\rangle$ value is also found for the SMC.
Interestingly, the mean UCAC2 LMC proper motion  agrees very well with
the  only work in  the  literature (Anguita  et al.  \cite{anguita00})
pointing to an unbound Magellanic Clouds-Milky Way interaction.
Nonetheless, the implications of  the UCAC2 proper  motion are hard to
reconcile with our  present   day understanding of  the  Clouds-Galaxy
interaction unless one assumes a more massive Milky Way.
Consequently, although  no   sources of  systematic  error have   been
identified, it is perhaps most likely that the UCAC2 catalog has an as
yet unidentified  systematic  error   resulting in  an   inconsistency
between UCAC2 and Hipparcos based results for the Magellanic Clouds.

 }

\authorrunning{Momany \& Zaggia}

\titlerunning{The LMC/SMC proper-motion}

\maketitle

\section{Introduction}
The  gas envelope  surrounding   both the  Large  and Small Magellanic
Clouds (LMC  and SMC) is reminiscent  of past interactions between the
two galaxies.   In turn,  the Magellanic  Stream  is evidence of  past
interactions   of   both galaxies   with the   Milky   Way (Kroupa  \&
Bastian~\cite{kroupa97}).   Modeling  the past  stellar and  dynamical
evolution of  the Magellanic  Clouds requires  good knowledge of  many
parameters,   in particular those   concerning their  spatial velocity
vector and  transverse velocity ($v_t$, van der Marel~\cite{marel04}).
A  sound determination of  the latter  parameter holds  the  key for a
better understanding of the Milky Way--Magellanic Clouds interaction.
The LMC/SMC transverse  velocity  can be  measured directly from   the
observed proper  motion of  the  galaxies, or  alternatively, via  (i)
dynamical modeling of  the Magellanic Stream  or (ii)  analysis of the
line-of-sight velocity field of the LMC/SMC.

Many groups have addressed the  $v_t$ measure of the Magellanic Clouds
via the direct method of  proper-motion (pm).   Being closer, the  LMC
has been  the subject of  more detailed studies (see Table~\ref{t_lit}
for a compilation of pm values).
At first sight,  one notes an overall  excellent agreement between the
first 6 determinations. However, there is one outlier ($\sim5.5\sigma$
times the  average  value  in  declination), namely  that  of Anguita,
Loyola and Pedreros (\cite{anguita00}, hereafter ALP00).
The ALP00 study is  based on 125 CCD  images spanning 8 years, and put
on a QSO absolute reference frame. It was the first pm analysis of LMC
stars reaching $V\simeq22$; i.e.  including  nearly all its  composite
stellar populations.

In a later work, Pedreros  et al.  (\cite{pedreros02}) using a similar
CCD data set and the same QSO method, contradicted their earlier ALP00
results and derived Hipparcos-{\em consistent} pm values.
Puzzled by   this discrepancy,   Pedreros et   al. (\cite{pedreros02})
searched for systematic errors possibly affecting the ALP00 study.
Their      most plausible explanation    pointed   to  peculiar pm  or
peculiarities in  the  rotation curve  of the  LMC  between  their two
studied fields (located at  different positions and distances from the
LMC center).
Further progress on this issue awaited results from  the 9 million LMC
stars of the MACHO project.  To our knowledge, the only MACHO study on
the LMC  pm is that    of Drake et al.  (\cite{drake02}),   confirming
``normal'' pm values.
Thus,  the  ALP00  study  remains the   only, inconsistent,   pm value
pointing to the Magellanic Clouds being {\em unbound}.\\
Following the methodology outlined in Momany et al.  (\cite{momany04})
we  analyze UCAC2  (Zacharias  et al. \cite{zach04})   pm  data of the
Magellanic Clouds.
Interestingly, we  find that the  UCAC2 data {\em unequivocally} point
toward a high $\langle\mu\delta\rangle$  of the LMC,  as in the  ALP00
study.
In  this paper we  attempt to assess the reliability  of  the UCAC2 pm
data by searching for  color/magnitude dependencies or other systematic
effects.  The UCAC2 pm data show no significant evidence of systematic
errors.
The  Hipparcos-UCAC2  discrepancy  basically  remains   unsolved.  The
plausibility and   implications of  the   UCAC2 and ALP00    values is
discussed.
%

%----------------------------------------------------------
\begin{table}[t]
\begin{center}
\caption{A compilation of LMC proper motion studies. }
\begin{tabular}{llll}  
\hline\hline
\noalign{\smallskip}
Source & $\mu \alpha$ & $\mu \delta$ & System \\
 &     (mas yr$^{-1}$) & (mas yr$^{-1}$) \\
\hline
Kroupa et al. (1994)              &  $+$1.30$\pm$0.60 & $+$1.10$\pm$0.70 & PPM\\
Jones et al. (1994)               &  $+$1.20$\pm$0.28 & $+$0.26$\pm$0.27 & Galaxies\\
Jones et al. (1994)               &  $+$1.37$\pm$0.28 & $-$0.18$\pm$0.27 & Galaxies\\
Kroupa Bastian (1997)$^{\dag}$    &  $+$1.94$\pm$0.29 & $-$0.14$\pm$0.36 & Hipparcos\\
Drake et al. (2002)$^{\dag}$      &  $+$1.40$\pm$0.40 & $+$0.38$\pm$0.25 & MACHO\\
Pedreros et al. (2002)$^{\dag}$   &  $+$2.00$\pm$0.20 & $+$0.40$\pm$0.20 & QSO\\
\hline 
Anguita et al. (2000)             &  $+$1.70$\pm$0.20 & $+$2.90$\pm$0.20 & QSO\\
\hline
\noalign{\smallskip}
\hline
\end{tabular} 
\label{t_lit}
\end{center}
%---------------------------------------------------------------------------------
\begin{list}{}{}
\item[$\vdag$] {The reported pm are those corrected and listed in
van der Marel~(\cite{marel02}).}
\end{list}%---------------------------------------------------------------------------------
\end{table}
%%%%%%%%%%%%%%%%%%%%%%%%%%%%%%%%%%%

\section{The UCAC2 LMC and SMC proper motion}
The  recent   availability of  large,  high precision  astrometric and
photometric catalogs like 2MASS   (Cutri et al.  \cite{cutri03}), SPM3
(Girard   et al.    \cite{girard04}) and   UCAC2  (Zacharias et    al.
\cite{zach04})  allow kinematic investigations of structures within
the  Galaxy  and determination of  the orbits  of its nearby satellite
companions.
In Momany et al. (\cite{momany04}), using the UCAC2 data, we have been
able to recover/confirm  the  HST pm  measurement of  the  Sagittarius
dwarf    galaxy   located    at      a    distance  of     25     kpc:
$\langle\mu\alpha,\mu\delta\rangle\simeq(-2.5,-2.0)$~mas/yr (Irwin  et
al. \cite{irwin96}).
This   value    has also  been   recently   confirmed  by  Dinescu et.
al. (\cite{dinescu04}) using the SPM3 catalog.
Thus, the Sagittarius dwarf is an excellent example of the reliability
of UCAC2 pm data, and  encourages the use of  these data for a similar
analysis for the Magellanic Clouds.

Combining the 2MASS infrared photometry with UCAC2 optical and pm data
offers    a unique  opportunity   to    discriminate LMC/SMC   stellar
populations from the Galactic (disk/halo) contribution.
On the one hand, this will  help identify possible dependencies of the
pm data on the employed photometric  magnitudes and colors, and on the
other,  it dramatically  expands the number  of  pm stars in different
LMC/SMC  populations  (having  different colors  and magnitudes). This
reduces the error budget  and possible contamination effects.  The net
result is a more robust estimate of the bulk pm.
Zacharias  et al.\ estimate  UCAC2 pm  errors to  be $1-3$  mas/yr for
12$^{th}$ magnitude stars and  about $4-7$ mas/yr at  16$^{th}$
magnitude.
Throughout this paper, we use absolute pm data for LMC and SMC stellar
populations having $12\le R\le16$.  Following a $3\sigma$ clipping, we
estimate the mean rms of the average pm  values to be $\le4.0$ mas/yr.
Although     quite significant, such      an  rms error  is acceptable
considering the  distance   of the  Magellanic  Clouds.   Moreover, we
prefer to  assume  such  a  high  error rather  than  that obtained by
dividing it by the square root of the number of stars.

%------------------------------------------one colonn figure
\begin{figure*}
%%--
%\centering \includegraphics[width=9cm,height=9cm]{FIG1.ps}
\centering \includegraphics[width=\textwidth]{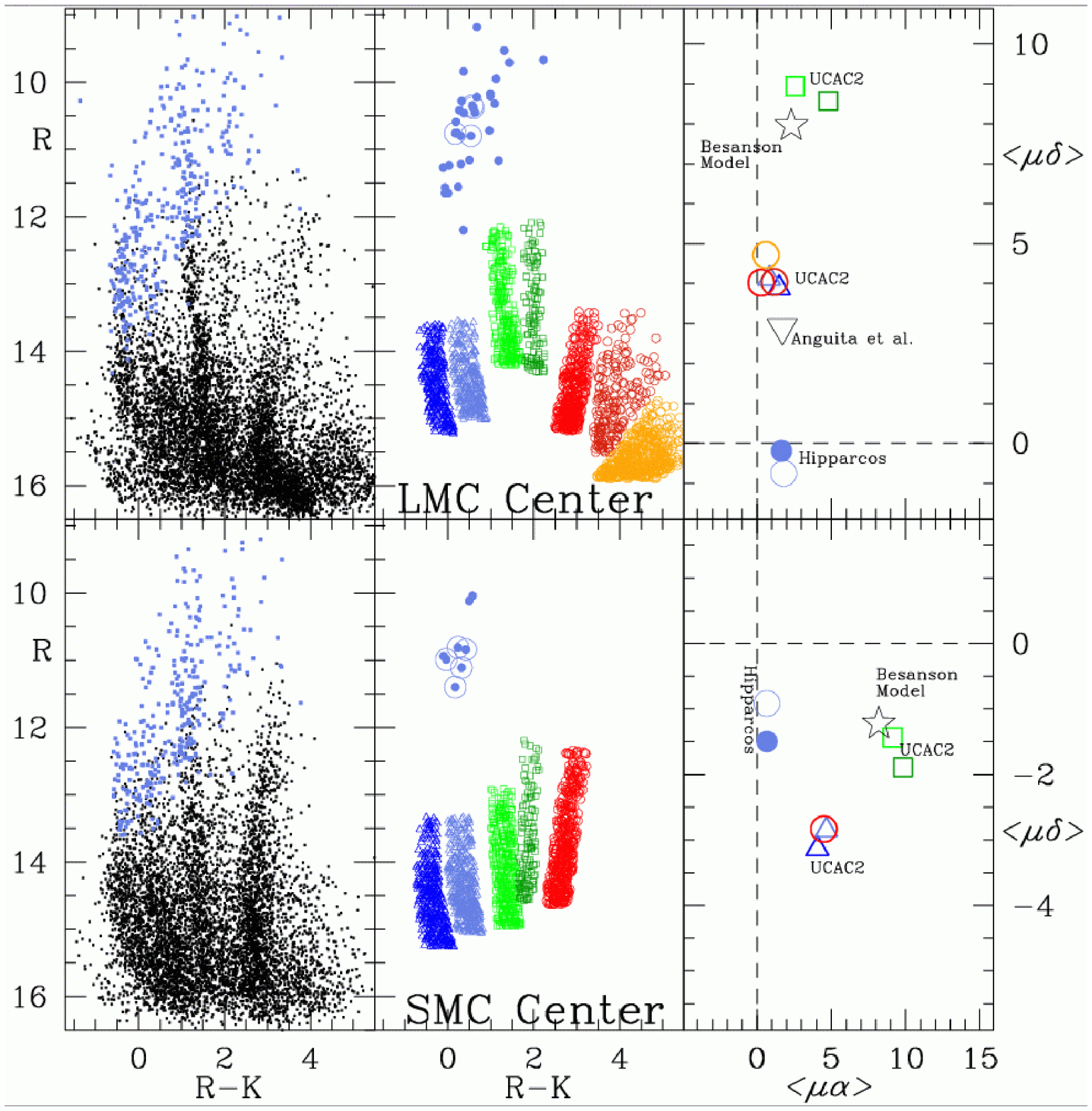}
\caption{
{\it  Left  panels}  display   UCAC2/2MASS  $R,(R-K)$  color-magnitude
diagrams of the LMC and SMC.
{\em Dark symbols}  are used for stars  whose pm originates {\it only}
from the AC2000.2 catalog, while {\em light symbols}  have a pm coming
from a number of catalogs, including Hipparcos.
{\it   Middle panels}  highlight  the  extracted populations:  the two
vertical sequences  around $(R-K)\simeq1.5$  are Galactic disk  stars,
the rest are LMC/SMC populations.
Dots brighter  than  $R\sim12.0$ are Hipparcos  samples  and tabled in
Kroupa \&  Bastian  (\cite{kroupa97}). Those plotted   as open circles
fall within the examined $1^{\circ}$ radius UCAC2 field.
{\it   Right panels}  show  the mean   proper  motion of the  selected
populations   in  the $\langle\mu\alpha,\mu\delta\rangle$    plane. The  different
symbols correspond to those in the middle panels. 
To these  we  added, an {\em  upside down  open  triangle} marking the
ALP00 value,  while an {\em open  starred symbols} shows  the expected
mean pm of Galactic disk stars (as derived from the
\besancon\ simulation).}
\label{f_fig1}
\end{figure*}
%-------------------------------------------------------------

The first epoch data for the pm come from different catalogs including
Hipparcos/Tycho, AC2000.2, as well   as re-measured AGK2, NPM  and SPM
plates. As in  Momany et al.~(\cite{momany04}),  we limit the UCAC2 pm
data to {\em only one} first epoch catalog  in order to limit internal
inhomogeneities, namely the AC2000.2 catalog (Urban et al.~\cite{U98},
\cite{U01}).  Based on the {\em Carte du Ciel} Astrographic plates and
measurements, the construction of  the AC2000.2 was performed scanning
all   the collected  plates and then   collecting  all the  positional
measurements in a homogeneous system based on Hipparcos. A key feature
of AC2000.2 is the  mean epoch of the positions  which range from 1890
to 1940. In the particular case of the  LMC the mean epoch of AC2000.2
is $\simeq1893.5$.  Comparing the LMC mean  epoch of AC2000.2 with the
observational epoch  of the  UCAC2  (1998.1$\div$2002.9) gives {\em  a
very large} epoch span for the pm measurements, therefore reducing the
pm errors.

Figure~\ref{f_fig1} summarizes our analysis in deriving the LMC (upper
panels) and SMC (lower panels) mean pm.
The {\it left panels} of Fig.~\ref{f_fig1} show  combined UCAC2/2MASS
$R,(R-K)$ color-magnitude diagrams, for two fields ($1^{\circ}$ radius
each) centered on the LMC and SMC.  {\em Dark} symbols are stars whose
pm  originates only from the AC2000.2  catalog (and used in estimating
the  pm),  whereas {\em light}  symbols  have   pm from  a  number  of
different catalogs, including Hipparcos.
The {\it middle    panels} highlight  the selected   LMC/SMC  stellar
populations.  Relying on the color-magnitude  diagrams in Nikoleav \&
Weinberg~(\cite{nikolaev00}) and  Alcock et al.~(\cite{alcock00}) of
the  LMC, one  can  easily disentangle   LMC stellar populations  from
the foreground Galactic contribution.  Indeed,  the two left-most selected
populations (open   triangles) are  almost  certainly LMC   young main
sequence  and  Helium burning stars.   On the  redder  side, the three
right-most populations (open circles) are red super-giants, oxygen- and
carbon-rich LMC stars, respectively.  In  between the LMC populations,
one identifies two    vertical  sequences (open squares)   which   are
Galactic disk F-K dwarfs and K giants.
As noted by  Alcock et al.~(\cite{alcock00}),  in these regions of the
color-magnitude diagram  the   Galactic disk  stars  out-number  those
belonging to both the Thick Disk and Halo combined.  A confirmation of
this comes from a \besancon\  synthetic color-magnitude diagram of the
Galactic  contribution: modeling in  both  the LMC and SMC  directions
clearly  showed that  virtually {\em  no  Galactic stars}  fall in the
Magellanic Clouds selected regions.
Filled  circles   are  35  Hipparcos  stars    tabled  in   Kroupa  \&
Bastian~(\cite{kroupa97}, their Table~5  excluding star number 26222).
Hipparcos stars  plotted as open circles  are those falling inside our
$1^{\circ}$ radius UCAC2 fields.\\
\indent  The {\it right  panels} show  the averaged  $3\sigma$ clipped
$\langle\mu\alpha,\mu\delta\rangle$    diagram    of   all    selected
populations.  For the LMC  (upper panel)  we easily  distinguish three
separate groups, namely:
{\em (i)} the Galactic populations characterized by a high
$\langle\mu\delta\rangle\simeq8.5$ mas/yr;
{\em (ii)} the LMC populations clearly clustering around
$\langle\mu\alpha,\mu\delta\rangle\simeq(0.9,3.9)$ mas/yr; and 
{\em      (iii)}      the    Hipparcos         measurements         at
$\langle\mu\alpha,\mu\delta\rangle\simeq(1.5,-0.3)$ mas/yr.
Likewise for the  SMC (lower  panel):
{\em (i)} the Galactic populations at
$\langle\mu\alpha,\mu\delta\rangle\simeq(9.0,-1.5)$ mas/yr;
{\em (ii)} the SMC populations  at
$\langle\mu\alpha,\mu\delta\rangle\simeq(4.5,-3.3)$ mas/yr; and
{\em (iii)} the Hipparcos measurements at
$\langle\mu\alpha,\mu\delta\rangle\simeq(1.0,-1.0)$ mas/yr.
The right panels of Fig.\ref{f_fig1}  show the discrepancy between the
Hipparcos     and  UCAC2      pm        mean values,  mainly        in
$\langle\mu\delta\rangle$  for the LMC and  in both directions for the
SMC.  Overall, the UCAC2   data  indicate a  mean  pm for  the LMC  of
$\langle\mu\alpha,\mu\delta\rangle\simeq(0.84,+4.32)$ mas/yr based  on
$\sim1620$ stars and $\simeq(4.44,-2.92)$  mas/yr based on  $\sim1200$
stars for the SMC.
Lastly, we searched for possible variations  of the mean pm across the
LMC.    However, for  all  considered   fields,  the  resulting pm  in
declination  was   always  larger than    Hipparcos-like  values.   In
particular, we   were able to   recover/confirm the ALP00  pm value of
their 2 fields  centered at $5^h\, 57^m$ and  $5^h\, 58^m$. Thus,  the
inconsistency of UCAC2 data is found all over the LMC.\\
\indent
One may be surprised  by the high pm  values obtained for the Galactic
populations.  In this regard, we note that the  derived pm of Galactic
populations is very well reproduced  by the \besancon\ Galactic  model
simulations.
The    mean model pm  was  computed   on the synthetic color-magnitude
diagrams for   the  two vertical  sequences   and  is plotted  as   an
open-starred symbol in both the right-most  panels.  We emphasize that
these are  model expectations, meant only  to test  the reliability of
UCAC2 values for the disk stars.  The agreement between UCAC2 and
\besancon\ however remains quite significant.

\section{The UCAC2-Hipparcos discrepancy}

Unveiling the  origin of  the UCAC2-Hipparcos  discrepancy is  quite a
difficult task   since  it may be  due  to  many different  sources of
systematic errors.   In the  following  we list   a number of possible
causes.

\begin{itemize}
\item  {\em  Magnitude/color  dependencies}:  The  LMC/SMC  UCAC2/2MASS
selected populations span $\sim5.0$ magnitudes  in ($R-K$) showing  no
appreciable dependence   on  color.   Furthermore,  dividing   the 5
selected  populations in magnitude bins  {\em  did not} result in any
appreciable variation of  the mean pm.   
Hence, although the presence of  residual color/magnitude terms in any
pm  analysis is  almost unavoidable  (Platais  et al.  \cite{plat03}),
these   {\em  seem not}  to  be   significant   enough to explain  the
discrepancy;

\item {\em Zone systematics}:  In Zacharias et al. (\cite{zach04}) pm
comparisons in the LMC field were computed for a sample of each of 200
stars.   This  star   by   star comparison yielded   a   difference of
$\langle\Delta\mu\alpha,\Delta\mu\delta\rangle=(+0.2,-1.0)$~mas   when
compared   to   the  average      value    of van    der   Marel    et
al. (\cite{marel02}).
This    is      clearly    not        enough    to     explain     the
$\langle\Delta\mu\delta\rangle\simeq4.0$~mas           Hipparcos-UCAC2
difference.  Zacharias et al.  also checked for systematic differences
in the  UCAC2 data as a  function of declination zone (their Table~8),
again not showing any significant systematic effect;

\item  {\em  Galactic  contamination  in the  Hipparcos  sample}:  The
Hipparcos sample  has been selected on  the the  basis of  radial
velocity.   These  are  bright, young O/B    LMC/SMC  members which on
Hipparcos color-magnitude diagrams   (see also   Fig.~1) are   clearly
limited to the brightest regions ($R\le12.00$), and  in a narrow color
range  $0\le(R-K)\le1.5$.    In  this   color-magnitude   region  some
contamination  by  nearby (within 1~kpc)  disk main  sequence stars is
predicted in the
\besancon\  photometric/kinematic  simulation.   However the  expected
radial velocity  of these  simulated  stars is  well below  $100$~km/s
implying that no contaminants {\em should be} present in the Hipparcos
LMC/SMC sample;

\item {\em Hipparcos O/B LMC stars selection effects}: The Hipparcos LMC
stars  ($V\ge10.5$)  are  actually far  below  the  100\% completeness
magnitude   limit  of Hipparcos  ($V\simeq9$),  and  are  located in a
color-magnitude range where the completeness   rapidly falls to zero
around $V\simeq13$.
Moreover,  these stars are projected on  the main body  of the LMC and
are surrounded by a rich and crowded  background that could lead to an
incorrect centroiding measurement;

\item  {\em O/B  spectral type  dependencies  in  Hipparcos}: 
A recent detailed analysis  by Schroeder et al.~(\cite{schr04}) of the
Hipparcos parallaxes of Galactic O/B  stars pointed out that these are
heavily erroneous {\em leading to absolute magnitude estimates up to 5
magnitudes fainter}.
On the other hand, Pan et al.  (2004) identified  a {\em duplicity} in
the  star Atlas affecting the  centroiding of Hipparcos, and suggested
this as  the  cause  of the   anomalous (10\%  smaller  than  accepted)
Hipparcos distance for the Pleiades.
%-------
A bias in the parallaxes of bright O/B stars  in the Pleiades has also
been confirmed recently by Soderblom et al.  (2004).   It is not clear
to us  how  wrong parallaxes can   affect proper motion determination,
however we note that Hipparcos gives {\em non-negative} parallaxes for
most of the LMC  stars.  This is much  larger than the {\em  expected}
value for the LMC.

\end{itemize}

In summary, we  are unable to find  a consistent explanation as to why
there   exists a   discrepancy  between the  Hipparcos    and UCAC2 pm
measurements.
Any a priori comparison of the two catalogs  would certainly favor the
UCAC2 as  the data source on which  to base a  pm determination of the
LMC. Indeed: {\em  (i)} the large number   statistics involved in  the
UCAC2 LMC/SMC measurements; {\em(ii)}  the large time baseline for the
pm measurements; {\em (iii)} the  reliability of UCAC2 in  determining
the pm for the Sagittarius dwarf; {\em (iv)} the excellent consistency
of the mean pm of different  LMC/SMC populations with different colors
and magnitudes; and {\em (v)} the good reproduction of  the mean pm of
the Galactic populations in  the LMC/SMC fields,  are all  evidence in
favor of  the  UCAC2 measurements  with  respect  to  the limited  (in
magnitude and spectral class) Hipparcos sample.

\section{Discussion: implications of the UCAC2 proper motion}

Serious problems arise when one replaces  the Hipparcos with the UCAC2
value for the LMC proper motion.  Indeed, the physical implications of
UCAC2/ALP00 pm   contrast  with our present-day understanding   of the
Local Group   in    general, and  the    Cloud-Galaxy interaction   in
particular.
All methods used  to measure the  Clouds' transverse  velocity ($v_t$)
{\em argue against} the UCAC2 pm values.
Applying eq. (55) of van der Marel et  al. (2002) for a Hipparcos-like
pm value, the resulting three-dimensional  space velocity for the  LMC
is $\overrightarrow{v}_{\rm  LMC}=(-56,-219,186)$ km/s,   while   the
corresponding  total,      tangential   and radial      velocities are
$(293,281,84)$ km/s, in the Galactocentric rest frame.
On the   other   hand,    applying    the   UCAC2 LMC     pm     gives
$\overrightarrow{v}_{\rm LMC}=(-974,-195,-113)$  km/s, with   the same
velocities being $(1001,971,241)$ km/s. Similarly, the UCAC2 pm of the
SMC  would imply $\overrightarrow{v}_{\rm SMC}=(-634,-974,475)$  km/s,
and $(1256,1251,103)$ km/s.
These UCAC2-based  values are  hard   to reconcile with  our   current
understanding of the:

\begin{itemize}

\item {\em LMC rotation}: the UCAC2 implied tangential velocity of the
LMC, 971 km/s, is so  large that, if  present, should  appear as in  a
prominent solid-body  rotation component  as  expected from the  large
projected dimensions in the sky of  the LMC.  There  is no evidence of
such a high component (see van der Marel \cite{marel02});

\item {\em Local Group}: in the Galactocentric rest frame, there is no
evidence in Local Group  or nearby groups of  galaxies of such extreme
velocities.   As recently shown  by  Karachentsev (\cite{kara05}), the
velocity dispersion of dwarf galaxies under the influence of the Milky
Way  is $\sigma_{vr}\simeq86$~km/s.   Similar values  of  the velocity
dispersion  are found in   other nearby  groups like Andromeda,  CenA,
M81/82 etc., indicating that no peculiar motions are present.  The LMC
radial velocity based on  UCAC2 is almost $\simeq3\times$ this typical
velocity dispersion, i.e. much higher than the escape velocity;

\item {\em Magellanic Stream}:  the UCAC2 LMC transverse velocity puts
it at $\sim90^{\circ}$ from  the  direction of the  Magellanic Stream.
Although we still do  not know why the stream  is starless, it is hard
to envisage how this relatively young feature survives at right angles
to the motion of the LMC (see Mastropietro et al.  \cite{pietro04} for
recent modeling);

\item {\em  Magellanic Bridge}: analysis of the  Magellanic Stream and
Bridge  shows  evidence  that  the  Bridge is  {\em  old}  (Putman  et
al. \cite{putman03}).  This strengthens the  idea that the LMC and SMC
have {\em  jointly} orbitted around the  Galaxy at least  for the past
Gyr or so.  However, this  observational evidence is hard to reconcile
with  the hypothetical UCAC2  scenario where  there exists  a relative
velocity between the LMC and SMC of the order of 1033 km/s.

\end{itemize}

A scenario with  the LMC traveling at  a high velocity (first passage)
in   the vicinity of  the Milky  Way seems also   to contradict recent
observational (Cole  et al.  \cite{cole05})  and theoretical (Bekki \&
Chiba  \cite{bekki05})  studies   that   successfully  reproduce  the
observed   morphology, dynamics  and  formation  histories of the  LMC
assuming it has been  in tidal interaction  with the SMC and the Milky
Way.
The only viable scenario that allows for the UCAC2/ALP00 LMC pm is one
in which   the  Milky  Way  is   more massive   (within  50~kpc)  than
$3\div4\times 10^{12}$M$_\odot$   as suggested by  M{\' e}ndez  et al.
(\cite{mendez99}).  Interestingly, the M{\' e}ndez et al.  modeling is
based on  a proper motion analysis of  30,000 stars from  the Southern
Proper-Motion (SPM) survey,  and is  consistent  with the  Anguita  et
al. pm value of the LMC.  Nevertheless, the M{\' e}ndez  et al.  Milky
Way total mass estimate remains {\em much  larger} than those reported
in recent studies, e.g.  Sakamoto,  Chiba \& Beers (\cite{saka03}) who
estimate  the  mass  of the Galaxy  (within   50~kpc) to be $5.5\times
10^{11}$M$_\odot$.

In conclusion, the  UCAC2 data may  have been valuable  in determining
the pm of the Sagittarius dwarf at 25~kpc  and in the direction of the
Canis Major  over-density at $\simeq8$~kpc, yet  the UCAC2 pm data are
{\em hard to   reconcile} with our  present  day understanding of  the
Cloud-Galaxy interaction.
Given   that our study  has failed  to  identify sources of systematic
errors, the UCAC2/Hipparcos inconsistency  remains unsolved and awaits
high-accuracy HST, GAIA and SIM proper motion measurements.

\begin{acknowledgements}
We are indebted to the anonymous  referee whose comments have improved
this  paper.  We thank  Giampaolo  Piotto and Luigi  Bedin for helpful
comments on the manuscript. YM acknowledges  financial support by MIUR
under the program PRIN2003.  ZS acknowledges financial support by MIUR
PRIN2002:  "Stellar  Populations  in  the  Local Group  as  a  Tool to
Understand Galaxy Formation and Evolution".

\end{acknowledgements}

\end{document}